\documentclass[conference,compsoc]{IEEEtran}
\ifCLASSINFOpdf
\usepackage{graphicx}
\else
\fi
\usepackage{booktabs}
\usepackage{multirow}
\usepackage{xcolor}
\usepackage{tikz}
\def\checkmark{\tikz\fill[scale=0.4](0,.35) -- (.25,0) -- (1,.7) -- (.25,.15) -- cycle;} 

\begin{document}
%
\title{A Domain-Specific Modeling Language \\ 
for Complex Event Processing Rules}

\author{\IEEEauthorblockN{Herbertt Diniz, Kiev Gama and Robson Fidalgo}
\IEEEauthorblockA{Informatics Center (CIn-UFPE)\\
Federal University of Pernambuco\\Recife-PE, Brazil\\
Email: \{hbmd, kiev, rdnf\}@cin.ufpe.br}
}


%


\maketitle


\begin{abstract}
Complex Event Processing (CEP) is one technique used to the handling data flows. It allows pre-establishing conditions through rules and firing events when certain patterns are found in the data flows. Because the rules for defining such patterns are expressed with specific languages, users of these technologies must understand the underlying expression syntax. To reduce the complexity of writing CEP rules, some researchers are employing Domain Specific Modeling Language (DSML) to provide modeling through visual tools. However, existing approaches are ignoring some user design techniques that facilitate usability.  Thus, resulting tools eventually has become more complexes for handling CEP than the conventional usage. Also, research on DSML tools targeting CEP does not present any evaluation around usability. This article proposes a DSML combined with visual notations techniques to create CEP rules with a more intuitive development model adapted for the non-expert user needs. The resulting tool was evaluated by non-expert users that were capable of easily creating CEP rules without prior knowledge of the underlying expression language. 
\end{abstract}

\begin{IEEEkeywords}
Complex Event Processing, Domain-Specific Modeling Language, Semiotics, Feature Model.
\end{IEEEkeywords}

%
\IEEEpeerreviewmaketitle

\section{Introduction}
Nowadays many systems (e.g., enterprise systems, social networks) and devices (e.g., the Internet of Things) are producing data massively. Many techniques and tools are being developed to store and analyze these massive amounts of data being generated. There is also need to handle the continuous data almost at the same time it is being received, with fast responses in a very low latency. This introduces other techniques and tools for the processing of continuous flows of data \cite{chen2014data}.  The concepts of timeliness and flow processing are justifying new classes of systems to handle these flows, with two emerging models competing today: data stream processing and complex event processing \cite{cugola2012processing}.  

In this article, we are particularly interested in Complex Event Processing (CEP). Through the usage of expression languages, CEP allows to capture data patterns in flows of information easily and to describe how rule engines can process data from these flows. It makes possible to trigger events out of complex relations from data flows. However, the fact of rules being expressed with query languages makes it a niche specialty. It becomes almost exclusive for specialists users, hindering the creation of solutions by a broader audience \cite{rozsnyai2007concepts}.

As observed by Roznyai et al. \cite{rozsnyai2007concepts}, there is a lack of tools that allow users to reconfigure a system easily or to refactor services and components in event-based systems such as the ones employing CEP. The Web-based interface approach of Chen et al. \cite{chen2014complex} for managing CEP engines and drawing rules with a flow-based approach, is an attempt to reduce the labor-consuming work of designing rules. However, one of the limitations of such approach is that it is coupled to particular CEP engines and technologies. Other efforts \cite{bruns2014conf, Taher2013346, Boubeta-Puig2014445} provide an abstraction layer, relying on Model-Driven Development (MDD) and Domain Specific Modeling Language (DSML) to decrease the complexity of using CEP. With this approach it is possible to create rules with graphical and high-level textual definitions that can be transformed into lower level languages, avoiding syntax or typing mistakes.

However, a major issue with existing DSML approaches for CEP is that many of these tools become more complex for their own use than the conventional use of an event query language \cite{white2008reducing}. They abstract the problem of written language, but they introduce a usability issue on the composition of the concepts and diagrammatic theory. It is clear that no design or usability approach addressed to end users is applied to consider more robust visual concepts. For instance, semiotics \cite{chandler1994semiotics}, which is the study of signs, is very close to the study of models since these are made of signs and symbols \cite{favre2005foundations}. What we found that the few existing DSML tools for CEP described in literature did not show any concerns with the creation of signs used in the tools neither present any validation with actual users. Therefore, user perception of the developer tools was not an important concern in existing research. 

In the work we present here, we describe our DSML approach in the creation of a tool that uses many semiotics-based principles \cite{moody2009physics} to enhance the expressiveness of the elements used for visually composing CEP rules. The resulting tool allows users to model rules through visual elements that represent higher level abstractions and to generate queries for different CEP engines. Our approach followed the three-step metamodeling process proposed by Brambilla et al. \cite{brambilla2012model}. We also employed Feature-Oriented Domain Analysis (FODA) \cite{kang1990feature} to help to model the language and a strong foundation \cite{moody2009physics} for visual representation to model the tool aiming to improve its usability and effectiveness. An empirical assessment was performed as part of the validation with actual users. The practical tests observed the success rate of non-specialists using our tool and an assessment through a \textit{Post-Study System Usability Questionnaire} (PSSUQ) \cite{lewis1995ibm}.

The major contributions that make our work different from existing literature is that: we provide a feature model describing the characteristics of the language; we apply semiotics principles to enhance the expressiveness of the visual language, and we provide concrete evidence of the visual tool being easy to use and intuitive. The remainder of this article is structured as follows. Section \ref{section-background} presents theoretical background to better understand the context and concepts used in ou work; section \ref{section-relatedwork} analyzes related work; section \ref{section-proposal} introduces the proposal of this work; section \ref{section-solution} details the solution we implemented, followed by section \ref{section-validation}, where is presented the assessment accomplished; and section \ref{section-conclusion} concludes the article.

\section{Background}
\label{section-background}
This section provides some contextualization on CEP as well as on semiotics applied to DSMLs, which are two of the main concepts behind the work presented in this paper.
\subsection{Complex Event Processing}
CEP uses an event-driven approach. It allows changes in state
to be monitored as they happen, letting applications respond more time-efficiently than a batch approaches \cite{etzion2010event}. CEP is appropriate when there is a need to perform real-time or quasi-real-time processing of incoming information flows to produce new knowledge.
It requires an expressive language to describe how incoming information
has to be processed, allowing to specifying complex relationships among the
information items that flow into the engine and are relevant to sinks. The CEP engine captures sequences of data involving complex ordering relationships, allowing to perform filtering, correlation, and aggregation of data \cite{cugola2012processing}.

CEP are specialists systems that support decision-making, where the specialized knowledge is encoded by experts on the domain \cite{vincent2010internet}. Furthermore, this category of system uses "rules" (or event patterns) to define if the established goals (conditions) was satisfied. Two important concepts in this context are presented in the next subsections: events and language for CEP.

\subsubsection{Events}

Bass \cite{bass2007internet} reinforces that an event is anything that happens or that might be considered an occurrence. For instance, a change of temperature, incoming messages or sensor readings. 
Therefore, the word \textit{event} is overloaded, defining one the one hand the current meaning, and on the other hand the meaning related to information processing. The context of each event will set the meaning or destination. This paper will consider that an event is a change in the state of an object, or over the time slice\cite{luckham2008power}.

\subsubsection{Languages for CEP}
Eckert and Bry\cite{eckert2009complex} point out that the requirements 
for an event query language might be defined by the following four 
aspects illustrated below. The examples queries are written in the Event Processing Language (EPL) \cite{etzion2010event}
, which defines CEP rules for 
Esper\footnote{http://www.espertech.com/} technology, being very similar to SQL database queries:

\begin{enumerate}
\item \textit{Data Extraction}: Events should have relevant data to decide if and how to react to them. The data access should be feasible regarding queries and capable of feeding other data sources, and should allow the creation of new events.

\texttt{select * from MyEvent.win:keepall()}

\item \textit{Composition}: It should be viable to join many individual events, so their occurrences combined over the time can produce a complex event.

\texttt{select fraud.accountNumber as accntNum,
fraud.warning as warn,
withdraw.amount as amount,
MAX(fraud.timestamp, withdraw.timestamp) as timestamp,
'withdrawlFraud' as desc
from FraudWarningEvent. win:keepall() as fraud,
WithdrawalEvent. win:keepall() as withdraw
where fraud.accountNumber = withdraw.accountNumber}

\item \textit{Temporal Relatioships}: Query events that imply in temporal expressions conditions should be happening at a specific time window or in a specific order. Other relationships between events, such as causality, should also be considered.

\texttt{select * from Withdrawal.win:time(10 sec ) where amount >= 200}

\item \textit{Accumulation}: Queries should involve the negation (lack) 
or aggregation of data events not caught in a stream. Since it is continuous flows, 
events could be correctly received when the stream ends. Thus, these queries need to be applied to a finite slice (a time slice or "window") of a stream, where their outcomes could be clearly well defined.

\texttt{select avg(price) from stockTickEvent.win:time(30 sec)}
\end{enumerate}

\subsection{Model-Driven Development}
The development of the proposed language was made using a  
\textit{Model-Driven Development} (MDD) approach \cite{Atkinson2003}. Typically, its implementation is (semi) automatically generated by metamodels \cite{brambilla2012model}. 
With MDD it is possible to follow the paradigm both addressing a General-Purpose Languages(GPLs), as well following the implementation of a domain-specific language. A modeling language is defined based on three basics concepts \cite{brambilla2012model}:

\begin{itemize}
\item{Abstract Syntax}: Describes the structural language and how the different primitives can be joined, apart from any specific representation or encoding.

\item{Concrete Syntax}: Describes the specific representations of the modeling language, covering encoding and visual appearance problems. The concrete syntax can be both textual as well graphic. If the syntax is visual, the modeling will result in one or more diagrams.

\item{Semantics}: Describes the meaning of the elements defined in the language and the meaning of the differents ways of combining them.
\end{itemize}



\subsection{Visual Notation Foundations}
Moody \cite{moody2009physics} claims that we have neither theory nor a systematic body of empirical evidence to guide us in the evaluation and construction of visual notation in Software Engineering.
So, we take that work, strongly based on semiotics, as our theoretical foundation to guide us in the creation of useful and effective visual representations. According to Moody, the anatomy of a good visual notation consists of:

\begin{table}[h]
\centering
\label{my-label}
\begin{tabular}{lll}
\begin{tabular}[c]{@{}l@{}}Visual Syntax \hspace{2 mm} = \\ (Concrete)\end{tabular} & \begin{tabular}[c]{@{}l@{}}Graphical Symbols  \hspace{3 mm}  +\\ (Visual Vocabulary)\end{tabular} & \begin{tabular}[c]{@{}l@{}}Compositional Rules\\ (Visual Grammar)\end{tabular}
\end{tabular}
\end{table}

A good notation should have cognitive effectiveness which defines the accuracy, ease and the speed with which the human mind can process a representation \cite{moody2009physics}. In order to provide a scientific basis for the assessment and design of visual notations, some base theories are needed, such as \textit{Descriptive Theory}  \cite{shannon1949university}. It points out how and why friendly visual notations can improve the capacity of communication, becoming a base to \textit{Prescriptive Theory} which define principles on the transformation of visual notations projects. The \textit{Prescriptive Theory} \cite{gregor2006nature} is the basis for the construction of the concrete syntax in our proposition. The main principles we employed in our work are presented as follows:

\noindent \textbf{Semiotic Clarity}: There must be a one-to-one correspondence between constructs and symbols. Otherwise, one or more types of anomalies can occur, such as symbol redundancy, symbol overload, symbol excess and symbol deficit.

\begin{figure}[h]
  \caption{Complexity management over the hierarchy into levels \cite{moody2009physics}}
  \centering
\label{fig:Gerencia-complexidade}  \includegraphics[width=0.8\linewidth]{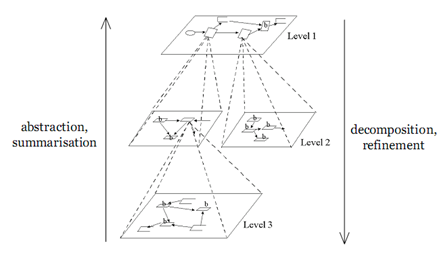}
\end{figure}

\noindent \textbf{Complexity Management}: proposes that a visual notation representing abstracted information should not generate overloading in the human mind. Figure \ref{fig:Gerencia-complexidade} presents the concept of complexity management, from utilization of levels hierarchy to define abstraction and decomposition of models \cite{moody2009physics}. Modularization is another good strategy for reducing the complexity of large systems.

\noindent \textbf{Graphic Economy}: It is related to the degree of complexity of notation, defined by the number of graphic symbols or size of the visual vocabulary \cite{nordbotten1999effect}. The use of more representative symbols generates \textit{graphic economy} by the fact that fewer symbols necessary to transmit the information, reducing the notation complexity.

\noindent \textbf{Cognitive Fit}: The notations should be done addressed to users, considering mainly the difficulties and the means that the diagrams will be transmitted (computer program, paper, blackboard, etc.).

\noindent \textbf{Perceptual Discriminability}: The accurate discrimination on how symbols differ from each other is paramount for appropriate interpretation of diagrams.

\section{Related Work}
\label{section-relatedwork}

This section presents a brief analysis of the projects that are most similar to our proposition of DSML for CEP. We point out their main characteristics and limitations. The choice was the result of a systematic mapping following the recommendations of the guidelines suggested by Kitchenham \cite{guidelines-2007}. Due to space constraints, the details of this selection process were left out of this paper but are available online\footnote{https://goo.gl/HdPYLf}. The following three articles were the most emphasized according to the criteria of mapping analysis:

The work of Bruns et al. \cite{bruns2014conf} present a domain specific language for event processing in machine-to-machine (M2M) platforms. It has a general vision of an event class architecture of three different types (Machine Events, Key Performance Indice Events, and Operation Events). Based on this structure they created a language following the essential concepts of the Object Management Group (OMG)\footnote{http://www.omg.com} recommendations. The language is textual and presents Text-Text (T2T) transformation, using Backus-Naur Form (BNF) and the Xtext tool\footnote{https://eclipse.org/Xtext/}, for the creation of abstract and concrete syntax of the language. It was focused on the solar energy industry domain, to increase the productivity of software development and to improve the quality of CEP systems. However, they do not provide a clear description of the metamodel or how it was developed. Moreover, the tool focuses on text-to-text transformations, which might be considered a disadvantage against visual approaches, since it provides a lower abstraction \cite{engelen2010integrating}.

Thaer et al. \cite{Taher2013346} propose a solution using a model-driven approach to creating automatic adapters of incompatible web services interfaces. This solution uses model transformation in a graphical editor, which models interfaces with high-level abstraction and generates the suitable code for its adapters. The generated code is specific to the target complex event processing engine. Their approach has four main features: service interface modeling, compatibility test, incompatibility detection, adapter generation, and CCL code generation and deployment. 
The tool allows validation, model-model transformation and code generation using an MDD approach. The Graphic User Interface (GUI) employs an activity diagram notation. They GUI was built using the Epsilon tool \footnote{http://www.eclipse.org/epsilon/}, for semantical validation the \textit{Epsilon Validation Language} (EVL) was used, and the \textit{Epsilon Generation Language} (EGL) to generate the source code. Despite the fact that the activity diagram is a common notation to many users, cases where a more complex rule definition is required would demand the utilization of a large number of activities. It leads to a very cluttered graphical description, which may result in difficulties in the modeling of larger systems \cite{white2008reducing}. Besides, since this type of diagram is very homogeneous, it becomes less intuitive in the usage of different elements involved in a CEP rule. Thus, it is limited in regards to the \textit{perceptual discriminability} \cite{moody2009physics} of the visual syntax.

Boubeta-puig et al. \cite{Boubeta-Puig2014445} propose a solution also using a model-driven approach for CEP domain specialists and non-specialists, focused in the construction of event patterns using a graphic tool. It proposes the easier creation of patterns and generation of automatic code. The tool allows validation and model-model transformation, and the code is generated using an MDD approach. The GUI construction employs a notation that allows other EPL languages to be extended. It also used EVL for source code generation. As a consequence of using many detailed concepts,  a diagrammatic visual language with high visual expressivity presents visual noise and low semiotic clarity \cite{moody2009physics}. Therefore, this approach reveals itself as little intuitive. Tasks with a large number of graphic resources have a confusing visual description.

None of the analyzed approaches provided any details on assessments or validation made with actual users. They also claim fewer lines of code, but this is natural with DSML, and more productivity, although no quantitative or qualitative data is presented. We also found that only two of them had a visual approach, but did not follow any guideline or foundation aiming to be intuitive or ease to use.

\section{Proposal}
\label{section-proposal}
\begin{figure}[h]
  \caption{Metamodeling process used in the DSML development \cite{brambilla2012model}}
\centering
\label{fig:Abordagem-Direcionada}  \includegraphics[width=0.7\linewidth]{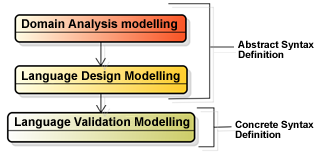}
\end{figure}

As analyzed in the literature, there is a major gap in the visual notation background of existed DSML tools for writing CEP rules. Here we attempt to address that by bringing a strong background in the modeling and conception of the solution, both in the metamodeling approach and in the modeling of the visual tool's elements.
The DSML developed in this work followed the three-step metamodeling process proposed by Brambilla et al. \cite{brambilla2012model} (Figure \ref{fig:Abordagem-Direcionada}). The first step consists on the \textit{Domain Analysis Modeling}, followed by the \textit{Language Design Modeling} where the construction of the abstract syntax happens, and finally the \textit{Language Validation Modeling} for the construction of the concrete syntax.

\subsection{\textbf{Domain Analysis Model}}
In this initial analysis we employed \textit{Feature-Oriented Domain Analysis} (FODA)\cite{kang1990feature}, which helped identify the main characteristics and the relationships
between the main components of the solution. The domain analysis was performed by analyzing the literature and CEP languages (Esper and Drools Fusion\footnote{http://drools.jboss.org/drools-fusion.html}), and interviewing CEP specialists.

\begin{figure}[h]
\caption{Feature model of a CEP rule}
\label{fig:Modelo-features}
\centering
\includegraphics[width=1\linewidth]{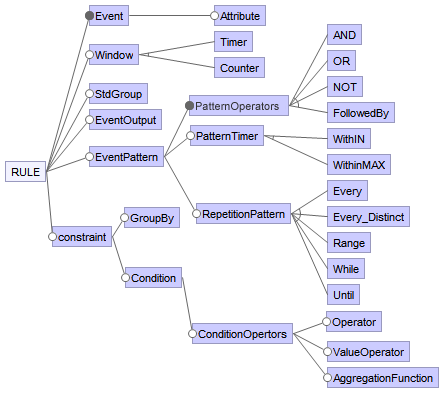}
\end{figure}

The feature model developed in this first phase was the output of the domain analysis, as illustrated in Figure \ref{fig:Modelo-features}.  It presents the necessary components to model and compose a CEP rule. In the following topics the description of each component defined in the feature model is presented:  

\begin{itemize}
\item \textit{Rule}: Represents a CEP rule. It has necessarily an \textit{Event}, and can optionnaly have a \textit{Window}, an event group (\textit{StdGroup}), an \textit{EventPattern}, an outcome event (\textit{EventOutput}) or a (\textit{Constraint});
\item \textit{Event}: They are the events of a CEP rule, with properties described through \textit{Attribute} components.
\item \textit{EventPattern}: Defines a logical composition for rules.
\item \textit{EventOutput}: Output of a triggered rule.
\item \textit{Window}: Represents the concept of a time window used in CEP rules. It can be temporal  (\textit {Timer}) or quantitative (\textit{Counter})
\item \textit{StdGroup}: It represents the concept of grouping, in simple events in the rules of CEP.
\item \textit{Constraint}: Logical restrictions that can be established in a CEP rule. They can be of two types: \textit{GroupBy} which represent constraints by grouping and \textit{Condition} type that set constraints through condition operators.
\item \textit{Attribute}: Defines event attributes or the structure of the elements of an event. 
\item \textit{Condition}: It has conditional operators: they can be a simple value (\textit{ValueOperator}), an attribute type operator (\textit{Operator}) or an aggregation function (\textit{Avg}, \textit{Sum}, \textit{Max}, etc).
\item \textit{PatternOperators}: Represents pattern operators, which can be \textit{AND}, \textit{OR}, \textit{NOT} or \textit{FollowedBy}, to define logical relationship between events. 
\item \textit{PatternTimer}: Defines temporal conditions in event patterns. They can be of type \textit{WithIN} and \textit{WithINMAX}.
\item \textit{RepetitionsPattern}: Defines repetitions patterns in programming, which can be of type \textit{Every}, \textit{Every\_Distinct}, \textit{Range}, \textit{While} and \textit{Until}.
\item \textit{ValueOperator}: Defines free values to compare with operators.
\item \textit{Operator}: Defines attributes to compare with logical operators.
\item \textit{AggregationFunction}: Defines aggregation fuctions of type: \textit{Avg}, \textit{Sum}, \textit{Max}, \textit{Min} and \textit{Count}, to compare with logical operators.
\end{itemize}

Besides being a typical output of the domain analysis, we chose the usage of a feature model because it is an approach with good expressivity and also able to remove ambiguities \cite{beuche2007software}. For instance, it is possible to express optional or mandatory components in a much cleaner way than in UML, which is the approach of other tools.

\subsection {\textbf {Language Design Modeling}}
The second step of the metamodeling process was broken down into two phases:
\begin{itemize}
\item\textit{Metamodel definition}:
After the feature model was defined, the metamodel was specified to refine the concepts of the feature model and to formalize the conceptual model, creating the necessary basis for its implementation.

\item \textit{Object Constraint Language}: This phase defines the basic constraints based on the defined model, preserving their integrity.
\end{itemize}

\subsection{\textbf{Language Validation Modeling}}

In the third step of the metamodeling process, the concrete syntax was built. The graphical representation of the metamodel was omitted to preserve space, but can be accessed online \footnote{https://goo.gl/D16esS}. The metamodel was created with the intention of facilitating the use of the tool, through the incorporation of semiotics principles (\textit{Complexity Management}, \textit{Graphic Economy}, \textit{Cognitive Fit}) \cite{moody2009physics} in two concepts:

\textbf{Modularization}: The concept of logical grouping developed in the language is based on the principle of Modularization, which establishes channels with greater representativity for variables to be defined in a compartmentalized way. It can be seen that clusters are decomposed according to their purpose in the structure of the language of event processing languages (e.g., Esper). Thus, the rule component (\textit{Rule}) has its scope defined into four logical groups:
\begin{itemize}
  \item \textit{Bring Group}: Represents all logic, to appear in a query of the CEP code language.

  \item \textit{Target}: Groups logic related to CEP events. In addition to the concept of modularization, \textit{Target}, is also present in the concept of complexity management through hierarchies, which is the key part in this composition. It is possible to group all logic related to events related to this component.

  \item \textit{ConditionGroup}: Groups the conditioning logic so that an event occurs or information is brought in according to what has been established.

  \item \textit{GroupbyCondition}: Grouping the conditioning logic, by grouping specific attributes, for an event to occur or for information to be brought in accordance to what has been established.
\end{itemize}

\textbf{Hierarchy}: The hierarchical diagramming feature concentrates the definition of the event logic (\textit{Events}), event patterns (\textit{EventPatterns}), event windows (\textit{Windows}), and groupings in events (\textit{StdGroup}). The hierarchical model structures the logic of action rules of the diagrammatic model. The Target component is the representation of the concept of "boxes-within-boxes" based on Simon\cite{simon1996sciences}, representing an abstraction of the components that bind to it.

\section{Solution}
\label{section-solution}

This section presents details of the solution we implemented. We also discuss many of the concepts taken from Moody's work \cite{moody2009physics} and applied in the construction of the DSML CEP tool. The tool is freely available for download on GitHub\footnote{https://github.com/herbertt/DSMLCEP}. The Graphic User Interface(GUI) was built with Epsilon\footnote{http://www.eclipse.org/epsilon/} tool, the validation of the language with \textit{Epsilon Validation Language} (EVL) and \textit{Epsilon Generation Language} (EGL) for code generation. 

\begin{figure}[h]
\caption{Example of model rule for a  Pluviometer}
\label{fig:modelo-regra}
\centering
\includegraphics[width=1\linewidth]{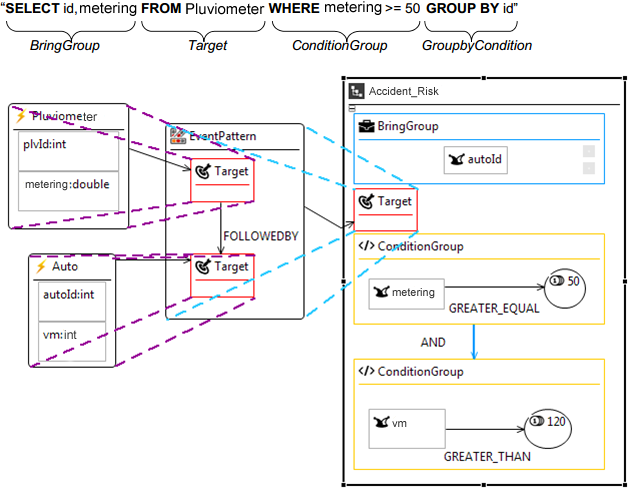}
\end{figure}

The tool allows creating graphically-defined rules to be converted to other CEP code languages automatically. Figure \ref{fig:modelo-regra} illustrates an example of the diagrammatic visual language showing a textual query in 
Esper's EPL and the corresponding visually modeled query. The current implementation can make transformations to EPL. We are currently implementing the transformation to Drools Fusion.

\subsection{Tool Overview}
\begin{figure}[h]
  \caption{General perspective of the DSML CEP tool}
\centering
\label{fig:general-view}  \includegraphics[width=0.8\linewidth]{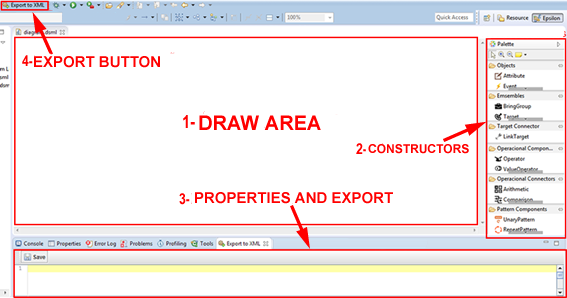}
\end{figure}

Figure \ref{fig:general-view} shows a general perspective on the tool. The screen is divided into draw area (1); constructors palette (2); properties and export (3); and export button (4). The \textbf{draw area} is where the user composes the visual diagrams representing the rules that originate the CEP queries. The \textbf{constructors} palette is discussed in more details on the next subsection since it involves many of the visual concepts from Moody \cite{moody2009physics}. In the \textbf{properties and export} tab it is possible to establish input values for instantiated component properties in the drawing area of the diagram, the export is responsible for displaying the text template transformation code. The save button saves the template and sends the rule to the message broker to deploy the generated query. The \textbf{export button} activates the Model-to-Text transformation (M2T), displaying results in the export tab.

\subsection{Visual Language Constructs}
In the constructor palette, described in Figure \ref{fig:general-view}, components are arranged so as to assemble the entire diagrammatic context. According to a logic of mnemonic association, we intentionally induce users to create their rules using components following an intuitive top-down sequence in the palette. The higher abstractions (e.g., objects) necessary for a rule are placed in the upper sections, and the more fine-grained details (e.g., operators) to build a rule are in the lower part of the palette. In this way, for any rule being modeled, a user starts naturally following this sequence of sections (top-down) to use the language constructs. It has symbols with a graphic representation, textual description and different formats to obtain a greater \textit{Graphic Economy} and \textit{Perceptual Discriminability}. The palette is organized into the following sections:

\noindent \textbf{Objects}: Establishes the main components of the drawing area.



\noindent \textbf{Ensembles}. Sets the grouping components, \textit{BringGroup}, \textit{ConditionGroup}, \textit{GroupbyCondition}, \textit{Target}, to segment the rule's structural logic in the concept of \textit{modularization} \cite{moody2009physics}.

\noindent \textbf{Target Connector}. Has a \textit{LinkTarget} component which is responsible for connecting to \textit{target} type groups. The dotted lines, depicted in Figure \ref{fig:Gerencia-complexidade} and in Figure \ref{fig:modelo-regra} attempt to represent a 3D vision of what a \textit{LinkTarget} represent in 2D, implementing the concept of \textit{Cognitive Fit} \cite{moody2009physics}.

\noindent \textbf{Operational Components}. Holds the operational components that perform the logic built into the groupings.

\noindent \textbf{Operational Connectors}. The  {Comparison, Arithmetic} and \textit{LogicalConnector} components represent the relationships of their binary logical types, between grouping components and operational components, except for the \textit{NOT} component.

\noindent \textbf{Pattern Components}. Holds the connectors and components which perform the logical operations built inside \textit{EventPattern}.

\begin{figure}[ht]
 \caption{Detail of the constructors sections.}
\centering
\label{fig:secao-objects} \includegraphics[width = 0.3 \linewidth]{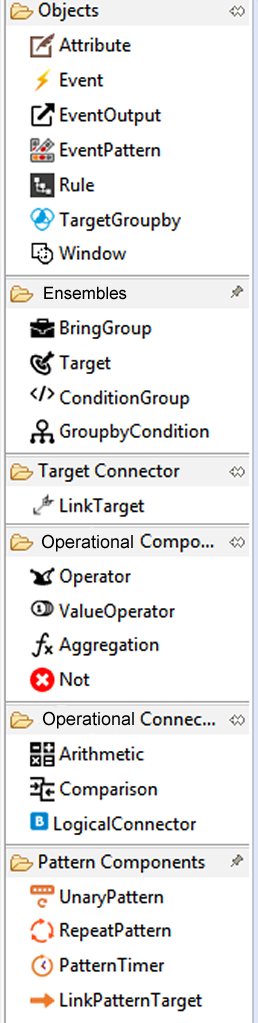}

\end{figure}


\section{Solution Assessment}
\label{section-validation}
Trying to address the lack of assessments or validation made with actual users in other DSML tools for CEP, we performed an experiment to evaluate the usability of the tool presented in this paper. This section presents the detailed assessment of the proposed solution in this work.

\subsection{Methodology}

The methodology we employed to evaluate the usability of the tool was based on: (1) the observation of the success rate of users visually modeling rules and generating correct queries; and (2) the analysis of the results of a Post-Study System Usability Questionnaire (PSSUQ) \cite{lewis1995ibm} applied to participants.

We ultimately intended to evaluate if the tool was \textit{easy} and \textit{intuitive}. The former concerns the solution allowing programmers not specialized in CEP to create rules patterns easily. The latter concerns the usability of the GUI, allowing any user, independent of their ability or programming knowledge, to be able to understand all the information and interact with the tool. We defined one-hour training, involving basic CEP content and usage of the tool, for leveling and presentation of content necessary to be used. The evaluation was organized as follows:

\begin{itemize}\item\textit{Experimental Group}: Nine participants (students from our university department) were involved in this experiment. Except for one participant who had previously worked with MDD and CEP in a different setting, the rest of the group had no previous experience with CEP, and only two of them had some prior experience with MDD. The data from a tenth participant was discarded to avoid bias because he was a beta tester that already knew the tool. 

\item\textit{Example Definition}: Two situations of CEP usage were describe as natural language in plain text in a generic domain (banking withdrawals). Participants had to read and understand the problem and visually model two rules (R1 and R2) using the tool.  

\item \textit{Solution Evaluation}: Following the two examples, to be graphically modeled. The participants of the experimental group executed each rule they defined comparing the result of its transformation with the respective correct responses in pre-defined queries written in EPL and made available to them only after both tasks being finished.
\end{itemize}

\subsection{Results and Discussion}
The outcomes of the assessment are summarized in Table \ref{table:evaluation-usability}, detailing the profile of each participant and which rules they successfully built using the tool:

\begin{itemize} \item \textit{1 PhD student}: Already worked with MDD tool and has used CEP tools sporadically. He did not complete the visual rules correctly.

\item \textit{4 BSc students}: Never worked with MDD tools and with CEP tools. All finished R1 correctly. In R2, two did not complete, and two completed correctly.

\item \textit {2 MSc students with MDD experience}: They had limited contact with MDD tools but have never used CEP tools. They completed all visual rules correctly.

\item \textit {2 MSc students without MDD experience}: Never worked with MDD and never used CEP tools. They completed all visual rules correctly.

\end{itemize}

\begin{table}[h]
\caption{Results and profiles of participating students}
\label{table:evaluation-usability}
\begin{tabular}{|p{5mm}|p{5mm}|p{5mm}|p{6mm}|p{5mm}|p{5mm}|p{6mm}|p{5mm}|p{5mm}|}
\hline
\multicolumn{1}{|l|}{\multirow{2}{*}{Level}} & \multicolumn{2}{c|}{Rules} & \multicolumn{3}{l|}{Experience in MDD} & \multicolumn{3}{l|}{Experience in CEP} \\ \cmidrule(l){2-9} 
\multicolumn{1}{|l|}{} & \multicolumn{1}{l|}{R1} & \multicolumn{1}{l|}{R2} & \multicolumn{1}{l|}{None} & \multicolumn{1}{l|}{Low} & \multicolumn{1}{l|}{High} & \multicolumn{1}{l|}{None} & \multicolumn{1}{l|}{Low} & \multicolumn{1}{l|}{High} \\ \hline

BSc. & \checkmark &  & \checkmark &  &  & \checkmark &  &  \\ 
BSc. & \checkmark &  & \checkmark &  &  & \checkmark &  &  \\ 
BSc. & \checkmark & \checkmark & \checkmark &  &  & \checkmark &  &  \\ 
BSc. & \checkmark & \checkmark & \checkmark &  &  & \checkmark &  &  \\ 
MSc. & \checkmark & \checkmark &  & \checkmark &  & \checkmark &  &  \\ 
MSc. & \checkmark & \checkmark &  & \checkmark &  & \checkmark &  &  \\ 
MSc. & \checkmark & \checkmark & \checkmark &  &  & \checkmark &  &  \\ 
MSc. & \checkmark & \checkmark &  &  &  & \checkmark &  &  \\ 
PhD. &  &  &  &  & \checkmark &  & \checkmark &  \\ 
\hline
\end{tabular}
\end{table}

As observed on Table \ref{table:evaluation-usability}, most participants did not have knowledge in MDD and CEP tools. Nevertheless, all of the eight subjects who are non-experienced with CEP could finish the first rule (R1) correctly. Among those, only two (25\%) of the non-experienced could not finish the second rule (R2). R2 is slightly more complex than R1. Therefore a smaller success rate was more likely. The surprising result was the experienced Ph.D. not being capable of finishing correctly. We suspect his previous knowledge may be a strong bias that affects his understanding of the tool. However, we have no concrete evidence to support that claim. 

Concerning the questionnaire, we merged two original PSSUQ questions into Q3 due to strong similarity of terms when translated into Portuguese (effectively versus efficiently). The questionnaire consists of a 1-7 Likert scale where one (1) stands for "Strongly Agree" and 7 for "Strongly Disagree." Grades greater than three (3), were considered as a good perception of the tool, four (4), was seen as neutral, and values greater than five (5) consisted of disapproval. When analyzing the Likert scale responses as intervals and applying descriptive statistics to it (Table \ref{pssuq-table}), we see a central tendency toward a positive perception of the tool. The only exception is Q11, which is a rather neutral response. Q18 provides an overall positive user perception of the tool. Therefore, the observation and the usability questionnaire results show evidence of good usability of the tool. Without much training, users could easily learn how to make CEP rules, intuitively and without help.

\begin{table*}[]
\centering
\caption{Statistics of responses (1-7 Likert scale) of the usability questionnaire \cite{lewis1995ibm}}
\label{pssuq-table}
\begin{tabular}{|l|r|r|r|r|}
\hline
\textbf{Question} & \textbf{Mean} & \textbf{St.Dev} & \textbf{Median} & \textbf{Mode} \\ \hline
Q1. Overall, I am satisfied with how easy it is to use this system       & 2,8           & 1,6            & 2               & 2             \\ \hline
Q2. It is simple to use this system        & 2,9           & 1,6            & 3               & 4             \\ \hline
Q3. I can effectively complete my work using this system        & 2,7           & 1,3            & 3               & 3             \\ \hline
Q4. I am able to complete my work quickly using this system       & 3,1           & 1,4            & 3               & 3             \\ \hline
Q5. I feel comfortable using this system       & 3,2           & 1,6            & 3               & 2             \\ \hline
Q6. It was easy to learn to use this system.       & 3,0           & 1,5            & 3               & 2             \\ \hline
Q7. I believe I became productive quickly using this system.       & 2,4           & 1,2            & 2               & 2             \\ \hline
Q8. The system gives error messages that clearly tell me how to fix problems.       & 3,9           & 2,1            & 3               & 3             \\ \hline
Q9. Whenever I make a mistake using the system, I recover easily and quickly.       & 2,7           & 1,3            & 3               & 3             \\ \hline
Q10. The information (on-line help, on-screen messages, documentation) provided with this system is clear      & 3,4           & 1,9            & 3               & 2             \\ \hline
Q11. It is easy to find the information I need      & 3,6           & 1,4            & 4               & 4             \\ \hline
Q12. The information provided with the system is easy to understand      & 3,3           & 1,6            & 3               & 3             \\ \hline
Q13. The information is effective in helping me complete my work      & 2,9           & 1,3            & 3               & 2             \\ \hline
Q14. The organization of information on the system screens is clear      & 2,4           & 1,7            & 2               & 1             \\ \hline
Q15. The interface of this system is pleasant      & 1,9           & 1,4            & 1               & 1             \\ \hline
Q16. I like using the interface of this system      & 2,2           & 1,3            & 2               & 2             \\ \hline
Q17. This system has all the functions and capabilities I expect it to have      & 3,1           & 1,8            & 3               & 4             \\ \hline
Q18. Overall, I am satisfied with this system      & 2,7           & 2,1            & 2               & 1             \\ \hline
\end{tabular}
\end{table*}

\subsection {Threats to Validity}
A significant point that can raise questions about the validity of the experiment is linked to the fact that the evaluation was done with students and not with professionals. However, according to Salman et al., \cite {salman2015students}, students and practitioners tend to perform similarly in software engineering experiments that evaluate development approaches in which participants do not have prior experience. Therefore, the use of only students in the analysis was considered little relevant. Another validity issue is how to objectively calculate if the solution developed is intuitive, the degree of intuition is purely subjective and can not be determined only by objective metrics. Also, the complexity and level of difficulty of the rules the participants had to model are limited. A scenario with more elaborate rules resulting in more advanced queries would be required for a deeper study. The empathy of some participants toward the authors of the tool may have influenced the responses. However, the subjects were instructed to be honest in their opinions.

\section{Conclusions and Future Work}
\label{section-conclusion}
Complex Event Processing is emerging as a critical approach to analyzing flows of information. It allows to capture patterns in such flows easily and to describe how rule engines can process flow data or fire events. However, there is a lack of tools supporting the easy construction of CEP rules. Existing DSML approaches for CEP trying to tackle that issue have cluttered metamodels for describing them and tools more complicated than the conventional use of an event query language. In part, this is due to a lack of foundation or guidelines in the construction of metamodels and the modeling tools. The lack of validation with actual users was another limitation found with existing approaches.

We followed a metamodeling process \cite{brambilla2012model} and a set of semiotics-based visual guidelines \cite{moody2009physics} for the construction of a CEP DSML tool that takes usability principles into account. The assessment we performed with non-experienced users show evidence of good usability of the tool. With little training, users could easily learn how to make CEP rules, intuitively and without help.

The major contributions are a feature model describing the characteristics of the language; the demonstration of semiotics principles to enhance the expressiveness of the visual language, and the concrete evidence of a visua DSML tool for CEP being easy to use and intuitive. As future work, we plan to do evaluations with more complex queries and integrate the tool into an event-based distributed platform. 



\bibliographystyle{IEEEtran}
\bibliography{IEEEabrv,DSML_CEP}

\begin{thebibliography}{10}
\providecommand{\url}[1]{#1}
\csname url@samestyle\endcsname
\providecommand{\newblock}{\relax}
\providecommand{\bibinfo}[2]{#2}
\providecommand{\BIBentrySTDinterwordspacing}{\spaceskip=0pt\relax}
\providecommand{\BIBentryALTinterwordstretchfactor}{4}
\providecommand{\BIBentryALTinterwordspacing}{\spaceskip=\fontdimen2\font plus
\BIBentryALTinterwordstretchfactor\fontdimen3\font minus
  \fontdimen4\font\relax}
\providecommand{\BIBforeignlanguage}[2]{{%
\expandafter\ifx\csname l@#1\endcsname\relax
\typeout{** WARNING: IEEEtran.bst: No hyphenation pattern has been}%
\typeout{** loaded for the language `#1'. Using the pattern for}%
\typeout{** the default language instead.}%
\else
\language=\csname l@#1\endcsname
\fi
#2}}
\providecommand{\BIBdecl}{\relax}
\BIBdecl

\bibitem{chen2014data}
C.~P. Chen and C.-Y. Zhang, ``Data-intensive applications, challenges,
  techniques and technologies: A survey on big data,'' \emph{Information
  Sciences}, vol. 275, pp. 314--347, 2014.

\bibitem{cugola2012processing}
G.~Cugola and A.~Margara, ``Processing flows of information: From data stream
  to complex event processing,'' \emph{ACM Computing Surveys (CSUR)}, vol.~44,
  no.~3, p.~15, 2012.

\bibitem{rozsnyai2007concepts}
S.~Rozsnyai, J.~Schiefer, and A.~Schatten, ``Concepts and models for typing
  events for event-based systems,'' in \emph{Proceedings of the 2007 inaugural
  international conference on Distributed event-based systems}.\hskip 1em plus
  0.5em minus 0.4em\relax ACM, 2007, pp. 62--70.

\bibitem{chen2014complex}
C.~Y. Chen, J.~H. Fu, P.-F. Wang, E.~Jou, and M.-W. Feng, ``Complex event
  processing for the internet of things and its applications,'' in \emph{2014
  IEEE International Conference on Automation Science and Engineering
  (CASE)}.\hskip 1em plus 0.5em minus 0.4em\relax IEEE, 2014, pp. 1144--1149.

\bibitem{bruns2014conf}
R.~Bruns, J.~Dunkel, S.~Lier, and H.~Masbruch, ``Ds-epl: domain-specific event
  processing language,'' in \emph{Proceedings of the 8th ACM International
  Conference on Distributed Event-Based Systems}.\hskip 1em plus 0.5em minus
  0.4em\relax ACM, 2014, pp. 83--94.

\bibitem{Taher2013346}
Y.~Taher, J.~Boubeta-Puig, W.-J. van~den Heuvel, G.~Ortiz, and I.~Medina-Bulo,
  ``A model-driven approach for web service adaptation using complex event
  processing,'' in \emph{European Conference on Service-Oriented and Cloud
  Computing}.\hskip 1em plus 0.5em minus 0.4em\relax Springer, 2013, pp.
  346--359.

\bibitem{Boubeta-Puig2014445}
J.~Boubeta-Puig, G.~Ortiz, and I.~Medina-Bulo, ``A model-driven approach for
  facilitating user-friendly design of complex event patterns,'' \emph{Expert
  Systems with Applications}, vol.~41, no.~2, pp. 445--456, 2014.

\bibitem{white2008reducing}
J.~White, D.~C. Schmidt, A.~Nechypurenko, and E.~Wuchner, ``Reducing the
  complexity of modeling large software systems,'' \emph{Designing
  Software-Intensive Systems: Methods and Principles: Methods and Principles},
  2008.

\bibitem{chandler1994semiotics}
D.~Chandler, ``Semiotics for beginners,'' 1994.

\bibitem{favre2005foundations}
J.-M. Favre, ``Foundations of model (driven)(reverse) engineering:
  models--episode i: stories of the fidus papyrus and of the solarus,'' in
  \emph{Dagstuhl Seminar Proceedings}.\hskip 1em plus 0.5em minus 0.4em\relax
  Schloss Dagstuhl-Leibniz-Zentrum f{\"u}r Informatik, 2005.

\bibitem{moody2009physics}
D.~Moody, ``{The “physics” of notations: toward a scientific basis for
  constructing visual notations in software engineering},'' \emph{IEEE
  Transactions on Software Engineering}, vol.~35, no.~6, pp. 756--779, 2009.

\bibitem{brambilla2012model}
M.~Brambilla, J.~Cabot, and M.~Wimmer, ``{Model-driven software engineering in
  practice},'' \emph{Synthesis Lectures on Software Engineering}, vol.~1,
  no.~1, pp. 1--182, 2012.

\bibitem{kang1990feature}
K.~C. Kang, S.~G. Cohen, J.~A. Hess, W.~E. Novak, and A.~S. Peterson,
  ``Feature-oriented domain analysis (foda) feasibility study,'' DTIC Document,
  Tech. Rep., 1990.

\bibitem{lewis1995ibm}
J.~R. Lewis, ``{IBM computer usability satisfaction questionnaires:
  psychometric evaluation and instructions for use},'' \emph{International
  Journal of Human-Computer Interaction}, vol.~7, no.~1, pp. 57--78, 1995.

\bibitem{etzion2010event}
O.~Etzion and P.~Niblett, \emph{Event processing in action}.\hskip 1em plus
  0.5em minus 0.4em\relax Manning Publications Co., 2010.

\bibitem{vincent2010internet}
\BIBentryALTinterwordspacing
P.~Vincent, ``{What is Complex Event Processing. The CEP Blog},'' 2010.
  [Online]. Available:
  \url{http://www.thetibcoblog.com/2010/03/12/the-return-of-the-expert-system
  (visited february 2, 2016)}
\BIBentrySTDinterwordspacing

\bibitem{bass2007internet}
\BIBentryALTinterwordspacing
T.~Bass, ``{What is Complex Event Processing. The CEP Blog},'' 2007. [Online].
  Available: \url{http://www.thecepblog.com (visited february 2, 2016)}
\BIBentrySTDinterwordspacing

\bibitem{luckham2008power}
D.~Luckham, ``{The power of events: An introduction to complex event processing
  in distributed enterprise systems},'' in \emph{International Workshop on
  Rules and Rule Markup Languages for the Semantic Web}.\hskip 1em plus 0.5em
  minus 0.4em\relax Springer, 2008, p.~3.

\bibitem{eckert2009complex}
M.~Eckert and F.~Bry, ``Complex event processing (cep),''
  \emph{Informatik-Spektrum}, vol.~32, no.~2, pp. 163--167, 2009.

\bibitem{Atkinson2003}
C.~Atkinson and T.~Kuhne, ``Model-driven development: a metamodeling
  foundation,'' \emph{IEEE software}, vol.~20, no.~5, pp. 36--41, 2003.

\bibitem{shannon1949university}
C.~Shannon and W.~Weaver, ``The mathematical theory of communication,''
  \emph{Urbana:Univ. of Illinois Press}, vol.~29, pp. 104--107, 1949.

\bibitem{gregor2006nature}
S.~Gregor, ``{The nature of theory in information systems},'' \emph{MIS
  quarterly}, pp. 611--642, 2006.

\bibitem{nordbotten1999effect}
J.~C. Nordbotten and M.~E. Crosby, ``The effect of graphic style on data model
  interpretation,'' \emph{Information Systems Journal}, vol.~9, no.~2, pp.
  139--155, 1999.

\bibitem{guidelines-2007}
B.~Kitchenham and S.~Charters, ``Guidelines for performing {S}ystematic
  {L}iterature {R}eviews in {S}oftware {E}ngineering,'' Keele Univ. and Durham
  Univ. Joint Report, Tech. Rep. EBSE 2007-001, 2007.

\bibitem{engelen2010integrating}
L.~Engelen and M.~van~den Brand, ``{Integrating textual and graphical modelling
  languages},'' \emph{Electronic Notes in Theoretical Computer Science}, vol.
  253, no.~7, pp. 105--120, 2010.

\bibitem{beuche2007software}
D.~Beuche and M.~Dalgarno, ``Software product line engineering with feature
  models,'' \emph{Overload Journal}, vol.~78, pp. 5--8, 2007.

\bibitem{simon1996sciences}
H.~A. Simon, \emph{{The sciences of the artificial}}.\hskip 1em plus 0.5em
  minus 0.4em\relax MIT press, 1996.

\bibitem{salman2015students}
I.~Salman, A.~T. Misirli, and N.~Juristo, ``Are students representatives of
  professionals in software engineering experiments?'' in \emph{Proceedings of
  the 37th International Conference on Software Engineering-Volume 1}.\hskip
  1em plus 0.5em minus 0.4em\relax IEEE Press, 2015, pp. 666--676.

\end{thebibliography}
%



\end{document}